\documentclass[english]{rspublic}
\usepackage{natbib}
\usepackage{hyperref}
\usepackage{bm}
\usepackage[T1]{fontenc}
\usepackage[latin9]{inputenc}
\usepackage{graphicx}
\usepackage{esint}
\usepackage{sidecap}
\makeatletter

\newcommand{\lyxdot}{.}

\makeatother

\usepackage{babel}

\begin{document}

\title [Quantum Criticality and Incipient Phase Separation...]
{Quantum Criticality and Incipient Phase Separation in the Thermodynamic 
Properties of the Hubbard Model}

\author[Galanakis, Khatami, Mikelsons, Macridin, Moreno, Browne, Jarrell]{
D. Galanakis$^{1}$, E. Khatami$^{2}$, K. Mikelsons$^{2}$, 
A. Macridin$^{3}$,
J. Moreno$^{1}$, D. A. Browne$^{1}$ and M. Jarrell$^{1}$}

\affiliation{$^{1}$Department of Physics and Astronomy, Louisiana State University, 
Baton Rouge, Louisiana, 70803, USA\\
$^{2}$Department of Physics, Georgetown University, Washington, District of Columbia, 
20057, USA\\
$^{3}$ Fermilab, P. O. Box 500, Batavia, Illinois, 60510, USA}
\maketitle

\begin{abstract}{Quantum criticality, DCA, Cluster methods}
Transport measurements on the cuprates suggest the presence of a quantum critical 
point hiding underneath the superconducting dome near optimal hole doping. We provide 
numerical evidence in support of this scenario via a dynamical cluster quantum Monte 
Carlo study of the extended two-dimensional Hubbard model. Single particle quantities, such as 
the spectral function, the quasiparticle weight and the entropy, display a crossover 
between two distinct ground states: a Fermi liquid at low filling and a non-Fermi 
liquid with a pseudogap at high filling. Both states are found to cross over to a 
marginal Fermi-liquid state at higher temperatures.
For finite next-nearest-neighbor hopping $t^\prime$ we find a classical critical 
point at temperature $T_c$. 
This classical critical point is found to be associated with a 
phase separation transition between a compressible Mott gas and an incompressible 
Mott liquid corresponding to the Fermi liquid and the pseudogap state, respectively.
Since the critical temperature $T_c$ extrapolates to zero as
$t^{\prime}$ vanishes, we conclude that a quantum critical point
connects the Fermi-liquid to the pseudogap region, and that the marginal-Fermi-liquid
behavior in its vicinity is the analogous of the supercritical region in the liquid-gas 
transition. 
\end{abstract}

\section{Introduction}

{\bf (a) Relevance of quantum criticality in the cuprates}

\begin{figure}[t]
\parbox[h]{6.5cm}{
\includegraphics[width=0.48\textwidth]{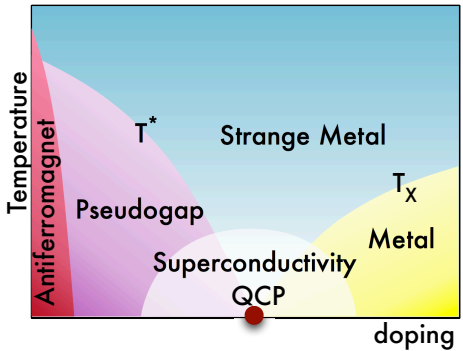}}
\parbox[h]{0.48\textwidth}{
\caption{The phase diagram of the cuprates. As a function of temperature and doping, the
cuprates display antiferromagnetic order at low doping, a non-Fermi liquid pseudogap region 
at intermediate doping and a metallic region at higher doping. Around optimal doping, 
superconductivity develops, and above the superconducting dome, a strange metal with non-Fermi
liquid properties appears. 
$T^*$ separates the pseudogap from the marginal Fermi-liquid phase. $T_X$ is the crossover 
temperature between the Fermi and the marginal Fermi-liquid regions. A quantum critical point
hides underneath the superconducting dome near optimal hole doping.}
\label{fig:crossover-phase-diagram-QCP}}
\end{figure}

The unusually high superconducting transition temperature of the hole doped 
cuprates~\citep{bednorz_muller} 
remains an unsolved puzzle, despite more than two decades of intense theoretical and 
experimental research. Pairing, which has a $d-$wave symmetry 
and short coherence length, but too high of a $T_c$ to be accounted by BCS~\citep{bcs}, 
is not the only unconventional property of these materials. Their phase diagram, shown 
in Fig.~\ref{fig:crossover-phase-diagram-QCP}, is a landscape of exotic states of matter. 
Undoped cuprates are Mott insulators with antiferromagnetic long-range order~\citep{neel_order}. 
Antiferromagnetism collapses upon small doping and it is replaced by a pseudogap state 
characterized by a suppression of spectral weight along the antinodal direction. Further 
doping turns the system into a conventional Fermi-liquid metal. Between the Fermi-liquid 
and the pseudogap region lies a strange metal phase with linear-$T$ resistivity. The 
superconducting dome emerges in the cross-over between the pseudogap and the Fermi-liquid 
regions at lower temperatures.

Strong electronic correlations are the cause of the rich phase diagram of cuprate 
superconductors~\citep{phillips_rmp}.
The same strong correlations render traditional theoretical approaches, such as perturbation 
theory and Fermi-liquid theory, inapplicable. Some recent conceptual progress has been 
achieved by associating the optimal $T_c$ with a  quantum critical point (QCP), lying 
underneath the superconducting dome and connecting the pseudogap and the Fermi-liquid 
regions~\citep{broun_criticality,sachdev_qcp}. Unlike a classical critical point, a QCP 
affects the behavior of the system in a wide range of temperatures and might explain the 
emergence of a linear-$T$ resistivity up to room temperature.

Experimental evidence for a QCP comes from transport~\citep{dirk,r_daou_09,f_Balakirev_09} 
and thermodynamic measurements \citep{tallon}. Angle-resolved photo-emission spectroscopy 
(ARPES) \citep{shen_nodal_quasiparticles,plate_overdoped_fermi_surface} and  quantum 
oscillation measurements \citep{DoironLeyraud_quantum_oscillations} show that in 
the pseudogap region, the Fermi surface consists of small pockets which have different 
topology than the large Fermi surface present in the Fermi liquid. It is reasonable to 
assume that those two states are orthogonal to one another and are connected through a 
transition. Additional evidence in support of quantum criticality comes from measurements 
of the Kerr signal in YBCO by Jing Xia {\em et al.}~\citep{Xia_KerrYBCO}. They find that 
at the pseudogap crossover temperature, $T^*$ a non-zero Kerr signal develops sharply and 
persists even inside the superconducting dome. This is consistent with earlier neutron 
scattering measurements by Fauqu\'e {\em et al.}\citep{Fauque_neutrons}, which show the 
development of magnetic order in the pseudogap phase.

In this manuscript we review numerical evidence of quantum criticality in the Hubbard model, 
the de-facto model for the cuprates, that appeared in earlier publications. In those cited works,
the Hubbard model is solved using the dynamical cluster approximation (DCA) in conjunction with 
several quantum Monte Carlo (QMC) cluster solvers. In all calculations relevant for the phase 
diagram we neglect the superconducting transition. The interplay
between the QCP and superconductivity will be discussed in a future publication.~\citep{Yang10}
In this review we focus on the thermodynamic quantities, such 
as the entropy and the chemical potential, and also on single-particle quantities, such as the 
spectral weight and the quasiparticle weight. The thermodynamic properties give unbiased evidence 
of quantum criticality, whereas single-particle properties may be used to gain more detailed insight 
on the ground state. Both set of quantities rely on the evaluation of the self-energy which can 
be calculated using quantum cluster methods.

At a critical interaction-dependent filling, we find that the entropy exhibits a maximum, 
the quasiparticle weight displays a crossover from Fermi liquid
to pseudogap behavior, and the spectral function shows a wide saddle point region crossing
the chemical potential. 
This is consistent with the presence of a QCP, since the lack of an energy scale results 
in an enhanced entropy at low temperatures. We also find that by tuning an appropriate 
control parameter, the next-nearest-neighbor hoping, $t^\prime$, the QCP becomes a classical 
critical point associated with a phase separation transition. We present our findings in two 
sections. In section \ref{section:fermiliquid_pseudogap}, we discuss the single-particle 
spectra and the thermodynamics properties of the $t^\prime=0$ Hubbard model. In section 
\ref{section:phase_separation}, we discuss the phase separation in the $t^\prime>0$ Hubbard 
model.

{\bf (b) Hubbard Model}

Short after the discovery of high-$T_c$ superconductors, Anderson~\citep{anderson} suggested 
that the Hubbard model captures the basic properties of the high temperature superconductors 
and Zhang and Rice~\citep{ZhangRice} demonstrated that only a single band is needed. 
The single-band Hubbard model is represented by the Hamiltonian:
\begin{equation}
H=-t \sum_{\left<i,j\right>,\sigma} \left[ c_{i\sigma}^\dagger c_{j\sigma}+\text{H.c.}\right]+
U \sum_i {n_{i\downarrow}n_{i\uparrow}},
\label{eq:hubbard_model}
\end{equation}
where $c_{i\sigma}^\dagger$ ($c_{i\sigma}$) is the creation (annihilation) operator of an 
electron at site $i$ and spin $\sigma$, $n_{i\sigma}$ is the corresponding number operator, 
$t$ is the hopping parameter between nearest-neighbor sites, and $U$ the on-site
Coulomb repulsion. Despite its apparent simplicity, the Hubbard model is notoriously difficult
to solve.  No analytical solutions exist except in one 
dimension~\citep{lieb_oned,frahm_oned,kawakami_oned}. However, tremendous theoretical and 
computational efforts have resulted in  approximation schemes that provide access to the physics 
of this model in higher dimensions. In this manuscript we also discuss results for the 
generalized Hubbard model which includes hopping between next-nearest neighbor with amplitude $t'$:
 \begin{equation}
H=-t \sum_{\left<i,j\right>,\sigma}\left[ c_{i\sigma}^\dagger c_{j\sigma}+\text{H.c.}\right]
-t' \sum_{\left<\left< i,l \right> \right>} \left[ c_{i\sigma}^\dagger c_{l\sigma}+\text{H.c.}\right]
+U \sum_i {n_{i\downarrow}n_{i\uparrow}}.
\label{eq:gen_hubbard}
\end{equation}

Important progress in our understanding of strongly correlated models has been achieved by the 
development of finite size methods, including exact diagonalization and QMC. The latter works 
well in the simulation of bosonic systems where creation and annihilation operators commute. 
However, due to the minus sign problem associated with the anticommutation relations of 
fermionic operators, QMC is limited to small lattice sizes and consequently give questionable 
predictions for correlated electronic systems in the thermodynamic limit. 

Another successful approach is the dynamical mean-field approximation (DMFA) which treats the 
local dynamical correlations explicitly and non-local (inter-site) correlations in a mean-field 
approximation~\citep{georges_dmft,metzner_vollhardt,e_mullerhartmann_89a,e_mullerhartmann_89b}. 
This technique becomes exact in the limit of infinite 
dimensions~\citep{georges_infinite_D,jarrell_infinite_D}. However, when applied to finite dimensions, the DMFA fails to describe the 
renormalization effects due to momentum-dependent modes and the transitions to phases with 
non-local order parameters. Thus, DMFA misses physical phenomena that are abundant in strongly 
correlated systems, such as the development of spin or charge density wave phases, localization 
in the presence of disorder, spin-liquid physics, unconventional superconductivity, etc. 

The limitations of the DMFA are addressed by cluster mean-field theories. Those fall into two 
categories~\citep{QCT}: the cluster dynamical mean field  theory (CDMFT)~\citep{kotliar_cdmft}, which is formulated 
in real space, and the DCA~\citep{hettler98} which is formulated in momentum space. In both cases the system is 
viewed as a cluster embedded in an effective medium. The formal difference between DCA and 
CDMFT is that in real space, the DCA cluster satisfies periodic boundary conditions whereas the 
CDMFT cluster is open. The two methods should give the same results for large enough clusters. 
In this work we present DCA~\citep{hettler98,hettler00} results. 

DCA treats short-ranged correlations explicitly, while longer ranged ones are approximated 
by the mean field. By increasing the cluster size, the length-scale of the explicitly treated 
correlations can be gradually increased while the calculation remains in the thermodynamic 
limit. In momentum space, the DCA can easily be conceptualized as the approximation in which 
the self-energy calculated by the coarse grained green function.  Quantum Monte Carlo based solvers such as Hirsch-Fye 
(HFQMC)~\citep{hirsch_fye}, continuous-time (CTQMC)~\citep{rubtsov_ctqmc} and 
determinantal quantum Monte Carlo (DQMC)~\citep{bss} are used to solve the cluster problem.
QMC methods are often formulated in imaginary time and an analytic continuation to real 
time is necessary to evaluate physical quantities. Fortunately, powerful techniques such as the 
maximum entropy method (MEM)~\citep{MEM,mem2} are able to successfully select the most likely 
solution.

Even though quantum cluster schemes have provided a tremendous breakthrough in our understanding 
of the Hubbard model, they are also subject to limitations. Quantum Monte Carlo solvers suffer 
from the sign problem, which scales exponentially with inverse temperature, interaction strength 
and cluster size. This limits the application of the method to relatively small cluster sizes, 
higher temperatures and intermediate interactions. The limitation in the cluster size is 
particularly problematic close to a phase transition where the correlation length diverges.
The coarse-graining also limits the momentum resolution, which for typical cluster sizes is 
too small to capture detail features of the spectra, such as van Hove singularities. 
For a Fermi liquid, this is not a limitation since 
the physics is dominated by the low frequencies in which the self-energy is momentum 
independent. However, intrinsically anisotropic states, such as the pseudogap,
or possibly the quantum critical region, can be captured only approximately.
Finally, MEM uses Bayesian statistics to find the most likely spectra for the QMC data, 
subject to sum rules, such as conservation of the spectral weight. Because of the 
statistical errors in the QMC data, the frequency spectrum resolved using MEM has a limited 
resolution.

Despite those limitations, progress can be achieved in accessing the quantum critical region by 
algorithmic optimizations. A truly universal way in dealing with the severity of 
the sign problem is to vastly increase the statistics, using massively parallel QMC algorithms
with highly optimized codes. 

\section{From Fermi Liquid to Pseudogap}\label{section:fermiliquid_pseudogap}

A great advantage of the DCA is its ability to evaluate the self-energy as a function of
momentum {\bf k} and Matsubara frequency $i\omega_n$, 
$\Sigma({\bf k},i\omega_n)$.  From the self-energy various single-particle quantities, 
such as the spectral function, $A({\bf k},\omega)$, the quasiparticle weight, $Z_{\bf k}$, 
and the energy can be derived. All those quantities provide insight about the ground state 
of the system. In this section we will show how the transition from the Fermi-liquid to the 
pseudogap state is reflected in such single-particle quantities.

{\bf (a) Spectral Function}

\begin{figure}
\includegraphics[width=\textwidth]{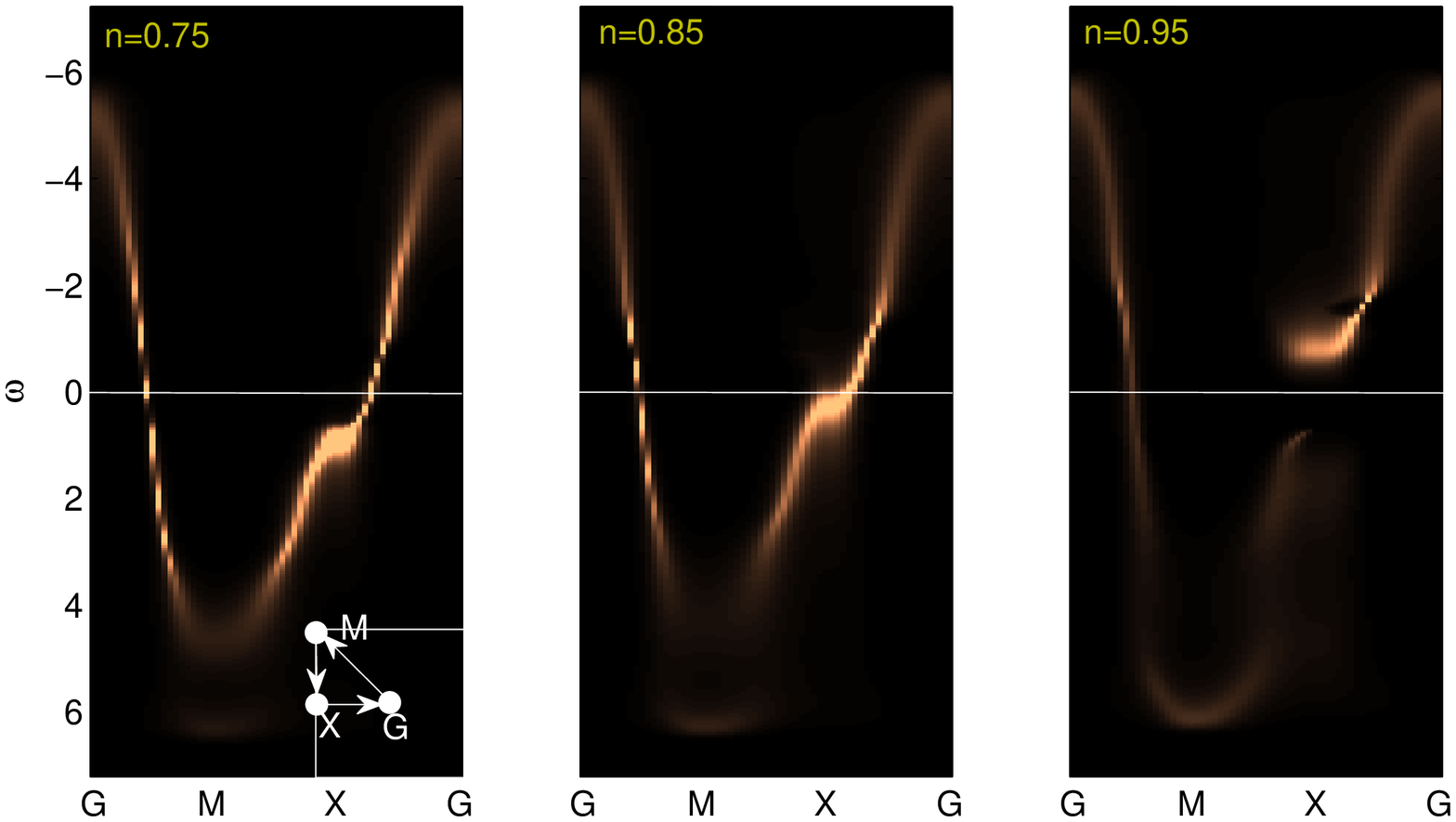}
\includegraphics[width=\textwidth]{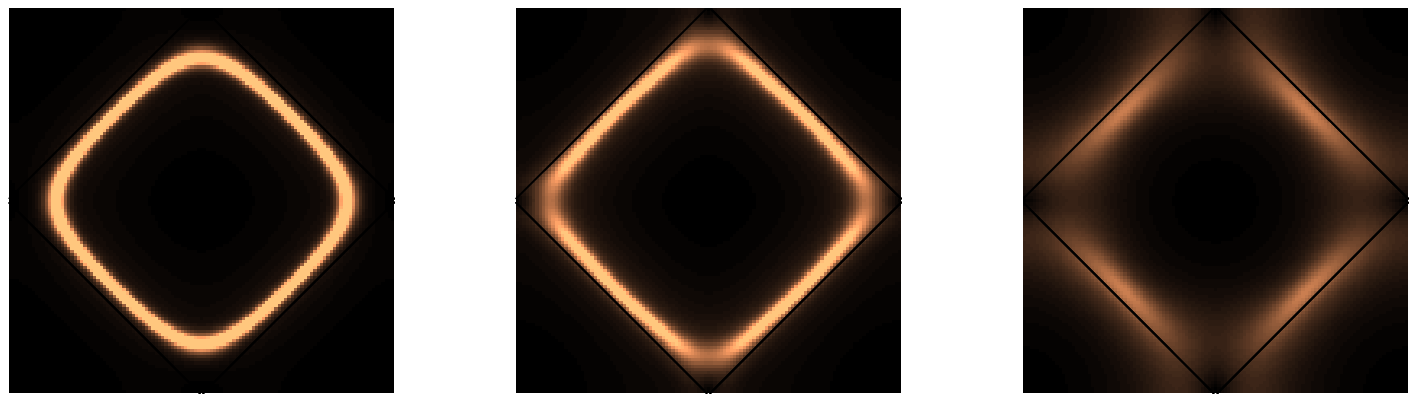}
\caption{Upper panels: Density plots of the spectral function $A({\bf k},\omega)$ for the Fermi 
liquid (left), marginal Fermi liquid (middle) and pseudogap region (right) for filling
$n=0.75, 0.85$ and $0.95$, respectively. (The dashed feature seen in the regions of steepest dispersion, 
especially for $n=0.75$, is a plotting artifact).
The momentum is along the path 
$G(0,0)\rightarrow M(\pi,\pi)\rightarrow X(\pi,0)\rightarrow G(0,0)$. A wide saddle point region 
between X and G sits above the chemical potential in the Fermi-liquid region and
crosses it around the critical filling ($n=0.85$). In the pseudogap region this features
sits below the chemical potential leaving a gap along the antinodal direction behind it.
Note that the fact that the dispersion looks discontinuous along $G(0,0)\rightarrow M(\pi,\pi)$ 
in the left and middle panels is an artifact of our interpolation algorithm.
Lower panels: Fermi surface as extracted from $|\nabla n_{\bf k}|$ in the Fermi liquid (left), 
marginal Fermi liquid (middle) and pseudogap (right) region showing the development of 
the pseudogap in the antinodal direction. The Coulomb repulsion is $U=6t$, the temperature 
$T=0.069t$, and the cluster size $N_c=16$. The energy unit is $4t$. 
\label{fig:single_spectrum}}
\end{figure}

The single-particle spectral function shows a clear evolution from a Fermi-liquid to a 
pseudogap state as the filling increases towards half filling. Fig. \ref{fig:single_spectrum} displays a density plot 
of the spectral function, $\displaystyle A({\bf k},\omega)=-\frac{1}{\pi}\Im G({\bf k},\omega)$, 
which is extracted by analytically  continuing the imaginary time Green function. 
At low filling, $n<0.85$, the spectral function exhibits a typical Fermi-liquid form. 
A notable characteristic is the presence of a wide saddle point region, reminiscent of a  
van Hove singularity,~\citep{Radtke94}  along the antinodal direction. Around the critical filling 
of $n=0.85$ this  saddle point feature crosses the chemical potential. 
This crossing results in a sharp peak in the density of states~\citep{raja_qcp},
which displays low-energy particle-hole symmetry~\citep{s_chakraborty_08}.
We are currently exploring the influence of the van Hove singularity on 
the superconducting transition.~\citep{Sandeep} 
At higher filling, $n>0.85$, the spectral weight collapses along the antinodal direction and 
a pseudogap opens. The Fermi surface
obtained by extremizing $\left|\nabla n_{\bf k}\right|$ shows a similar evolution (see lower
panels in Fig. \ref{fig:single_spectrum}). The 
Fermi-liquid region consists of a large hole pocket, which extends and touches the edges of the 
Brillouin zone $(0,\pm \pi)$, $(\pm \pi,0)$  at $n=0.85$. In the pseudogap region the Fermi 
surface consists of four Fermi arcs centered 
around the nodal points, similar to the ones obtained from ARPES. 
These results clearly demonstrate that the DCA can capture 
qualitatively the evolution of the ground state from a Fermi-liquid to a pseudogap phase.

{\bf (b) Quasiparticle Weight}

\begin{figure}
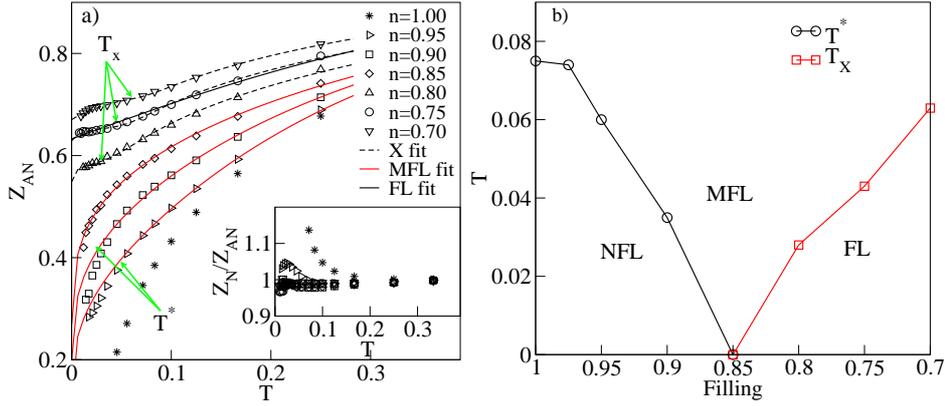

\begin{center}
\includegraphics[height=0.42\textwidth]{raja_quasiparticle_weight}
\includegraphics[height=0.42\textwidth]{temperatures}
\caption{{\bf a)} The antinodal quasiparticle fraction $Z_{AN}$ as a function of temperature 
for different values of filling, $U=6t$ and cluster size $N_c=16$ (the unit of
energy is $4t$). The onset of the pseudogap region is determined by the vanishing of the 
antinodal spectral weight at zero temperature. The dashed and solid lines represent fits of 
the low temperature ($T<0.3$) data to marginal Fermi liquid (red solid curves), Fermi liquid 
(black solid curves) and crossover forms (dashed black curves), respectively. The arrows show 
the corresponding crossover temperatures $T_X$ and $T^*$. The value of $T^*$ presented here
is obtained from the spin susceptibility as explained in~\citep{raja_qcp}, but is consistent
with the one extracted from the from the fitting forms. The ratio $Z_{N}/Z_{AN}$ of the 
quasiparticle weight in the nodal ($(\pi,\pi)$) and antinodal 
($(0,\pi)$) directions (inset) diverges as the pseudogap develops in accordance with 
Fig.~\ref{fig:single_spectrum}. {\bf b)} The crossover temperatures 
$T_X$ and $T^*$ as a function of filling as extracted from the temperature dependence of 
$Z_{AN}$~\citep{raja_qcp} for the same parameters.
\label{fig:Quasiparticle-weight-raja}}
\end{center}
\end{figure}

Whereas the spectral function gives a qualitative understanding of the ground 
state, it relies on the analytic continuation of numerical data. Since 
extracting quantitative information from analytically continued data is 
difficult, a more robust way is to rely on imaginary time quantities, such as the 
quasiparticle weight $Z({\bf k})$. Since the quasiparticle weight is 
finite across a Fermi surface, but  vanishes if the spectrum is incoherent,
it will allows to clearly distinguish between a Fermi liquid and a pseudogap state. 
The quasiparticle weight can be directly obtained from the Matsubara frequency self-energy as 
$\displaystyle Z_{0}\left({\bf k}\right)=
\left(1-\frac{\Im\Sigma\left({\bf k},i \omega_0 \right)}{\omega_0}\right)^{-1}$,
where $\omega_0=\pi T$ is the lowest fermionic Matsubara frequency. At the limit $T\rightarrow0$ 
and for a well-behaved self-energy, $Z_{0}\left({\bf k}\right)$ converges to the quasiparticle 
weight, $Z\left({\bf k}\right)$. Fig.~\ref{fig:Quasiparticle-weight-raja} (a) displays 
$Z_{AN}=Z_0(\omega_0=\pi T, k \parallel (0,0) \rightarrow (0,\pi))$, the Matsubara quasiparticle 
weight along the antinodal momentum direction for $U=6t$ and a cluster of size 
$N_c=16$~\citep{raja_qcp}. The momentum ${\bf k}$ at the Fermi surface is determined 
by maximizing $\left| \nabla n(\bf k) \right|$. $Z_{AN}$ exhibits two distinguishable 
behaviors: for $n>n_{c}=0.85$ the quasiparticle weight vanishes, whereas it approaches a 
finite value for  $n<n_{c}$.  The $n>n_{c}$ region corresponds 
to the pseudogap state in which the spectral weight collapses along the antinodal direction, 
while the $n<n_{c}$ region behaves as a Fermi liquid. 

The temperature dependence of $Z_{AN}$ 
(Fig. \ref{fig:Quasiparticle-weight-raja} (a)) not only provides information about the ground 
state but also allows the extraction of relevant energy scales. By comparing the numerical 
results with analytical expressions derived from particular phenomenological forms of the 
self-energy, we obtain $T_{X}$ and $T^*$. At low filling, $n<n_{c}$, the high $T$ dependence 
of $Z_{AN}$ is best fit by a marginal Fermi-liquid form, whereas for low $T$ 
the data is best fit by a Fermi liquid. The crossover occurs at a temperature $T_X$, which is 
extracted by fitting with a crossover function, and is accompanied by a change in the sign of 
the curvature of $Z_{AN}$. At higher filling ($n>0.85$), the high temperature     
$Z_{AN}$ can also be fit by a marginal Fermi liquid, whereas at low temperatures,
it cannot. The crossover temperature $T^*$ can be extracted as the lowest temperature where the 
marginal Fermi liquid fit lies within the statistical error. However a more accurate value can
be obtained from the bulk spin susceptibility which exhibits a peak at $T^*$ and the two values 
are found to be consistent~\citep{raja_qcp}. The crossover temperatures $T_{X}$ 
and $T^{*}$ are shown in Fig. \ref{fig:Quasiparticle-weight-raja} (b). Both of them converge 
to zero as the filling approaches $n_c=0.85$, which is the same value for which the peak in 
the density of states~\citep{raja_qcp} crosses the chemical potential. 

{\bf (c) Thermodynamics}

\begin{figure}
\begin{center}
\includegraphics[width=0.9\textwidth]{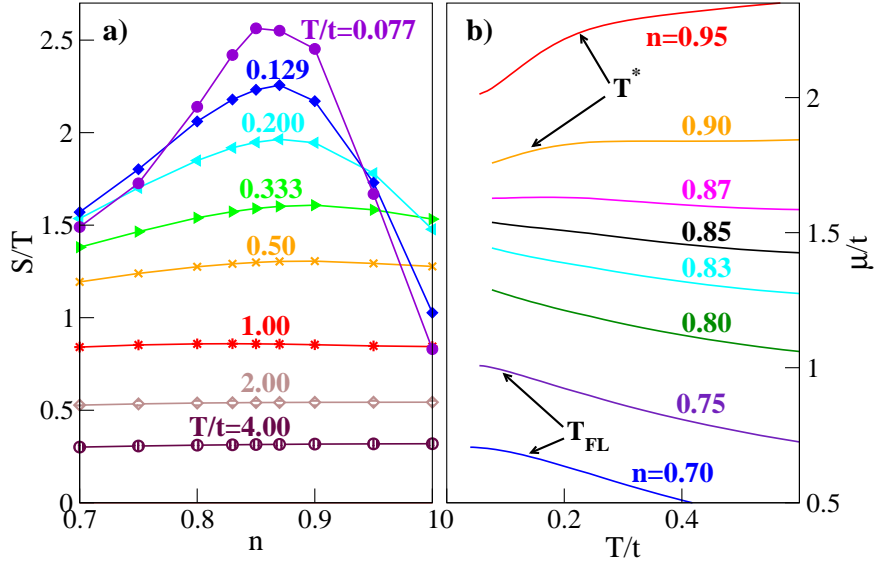}
\end{center}
\caption{{\bf a)} The filling dependence of the entropy divided by temperature, $S/T$, for 
various temperatures at $U=6t$ and $N_c=16$. With decreasing temperature a peak develops 
around the critical filling of $n_c=0.85$. {\bf b)} The temperature dependence of the chemical 
potential $\mu$ for different fillings. At the critical filling, $n_c$, $\mu$ becomes temperature 
independent at low temperatures.
\label{fig:Entropy-versus-filling}}
\end{figure}

A different perspective at the transition from a Fermi liquid to the pseudogap state comes from 
the evaluation of the entropy. We obtain the entropy by integrating the energy using 
the formula:
\begin{equation}
S(\beta,n)=S(0,n)+\beta E(\beta,n)-\int_{0}^{\beta}E(\beta^{\prime},n)d\beta^{\prime},
\label{eq:entropy_integral_energy}
\end{equation}
where $\beta$ is inverse temperature and $S(0,n)$ is the infinite temperature entropy. 
Equation \ref{eq:entropy_integral_energy} is appropriate for QMC calculations, because 
the integration reduces the statistical error. The challenge is to have good enough 
statistics to control the error of the surface term, $\beta E(\beta,n)$. In Mikelsons 
{\em et al.}~\citep{mikelsons_thermodynamics} large statistics was possible simply by
using large computational resources. The entropy divided by the temperature, shown in 
Fig. \ref{fig:Entropy-versus-filling} (a), exhibits a maximum at exactly the same critical filling 
that was identified before from the spectral function and the quasiparticle weight. In 
Fig.~\ref{fig:Entropy-versus-filling} (b), we show the chemical potential, $\mu$, as a function 
of temperature. We note that at the critical filling $d\mu/dT=0$,  since the entropy and the 
chemical potential are related by the Maxwell relation:
\begin{equation}
\left(\frac{\partial S}{\partial n}\right)_{T,U}=
-\left(\frac{\partial\mu}{\partial T}\right)_{U,n}.\label{maxwell_S_mu}
\end{equation}

\begin{figure}
\includegraphics[height=0.4\textwidth]{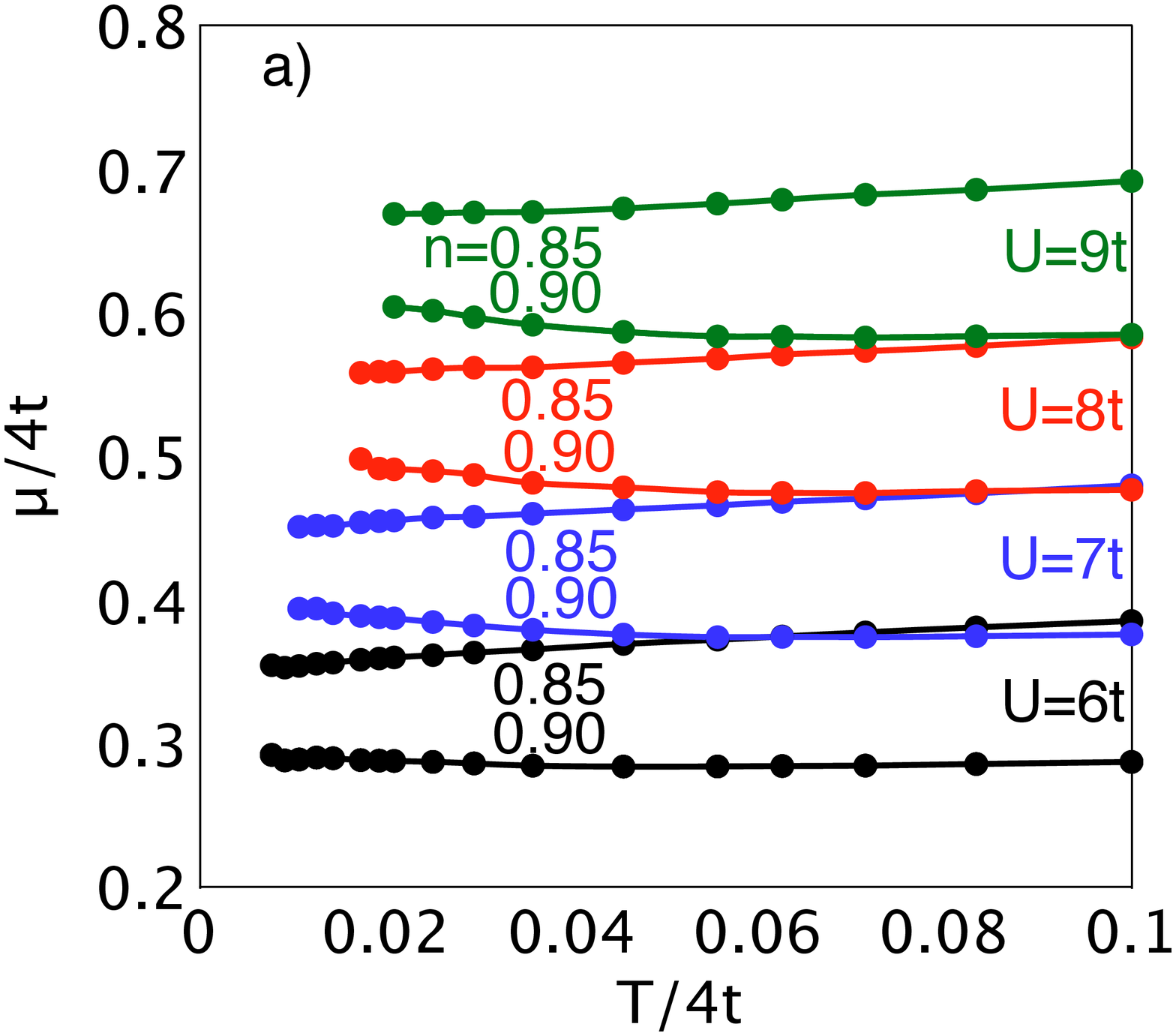}
\includegraphics[height=0.4\textwidth]{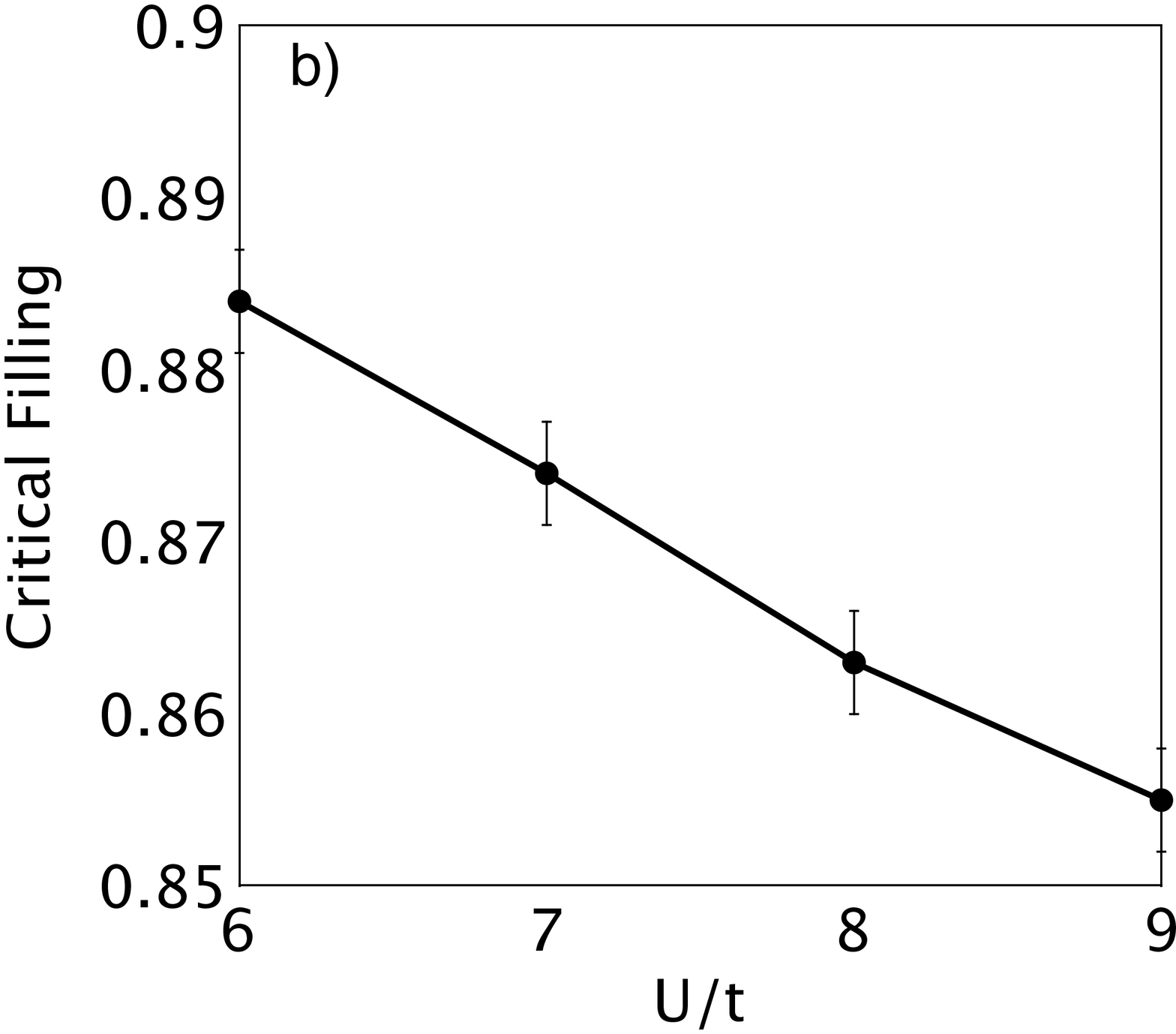}
\caption{{\bf a)} The chemical potential as a function of temperature for fillings of 
$n=0.85$ and $0.90$ and for a variety of interaction strengths $U$ for $N_c=12$. 
{\bf b)} The critical filling, defined by the filling in which $\partial\mu /\partial T=0$ 
versus $U$. The critical filling decreases with $U$ monotonically and is projected to reach 
the atomic limit value of $n_c=2/3$ at $U_c=30t$.\label{fig:nc_vs_U}}
\end{figure}

Also the temperature dependence of the chemical potential
 can be used as a practical criterion to identify the location of the critical 
filling, because evaluating the chemical potential is much less computationally intensive 
than evaluating the entropy. Using this criterion we investigate the important question 
of the dependence of $n_c$ on the Coulomb repulsion $U$. As it is shown in 
Fig.~\ref{fig:nc_vs_U}, we find that increasing $U$ reduces the critical filling
and thus enlarges the pseudogap region in the phase diagram. Our  
results follow the trend proposed in earlier arguments~\citep{s_chakraborty_08} according to which
the critical filling decreases in order to reach the atomic limit value of $n_c=2/3$.

In this section we have shown that several single-particle quantities are consistent 
with the presence of a QCP. The qualitative form of the single-particle  spectrum shown in 
Fig.~\ref{fig:single_spectrum} is fundamentally different in the Fermi-liquid and the pseudogap 
regions, which points to orthogonal ground states. The temperature dependence of the quasiparticle
weight reveals the presence of two crossover temperatures $T^*$ and $T_X$, which converge to 
zero at $n_c$ as shown in Fig.~\ref{fig:Quasiparticle-weight-raja} (b). If the crossover 
temperatures $T_X$ and $T^*$ constitute energy scales that suppress degrees of freedom, their 
vanishing at $n_c$ means that there are no relevant energy scales to quench the entropy and 
therefore it collapses at a slower rate, which is consistent with the peak of the entropy 
observed at $n_c$. The natural next step to investigate quantum criticality is to access the QCP. 
However the fermion sign problem severely limits the applicability of quantum Monte Carlo 
techniques close to a QCP.  It is possible however, as we will discuss in the next section, that 
by tuning an appropriate control parameter, the critical point may be lifted to finite temperature 
and thus studied with QMC.

\section{Phase Separation and Quantum Criticality}\label{section:phase_separation}
Experiments suggest that cuprate superconductors are susceptible to
charge inhomogeneities, such as stripes or checkerboard modulations~\citep{hinkov04}.
These inhomogeneous charge patterns have stimulated intense theoretical and experimental
research. Here we will consider the possibility that those charge instabilities
are evidence that the cuprates are close to a phase separation transition,
and this proximity may be related to the nature of the QCP.

\begin{figure}
\begin{center}
\includegraphics[height=0.40\textwidth]{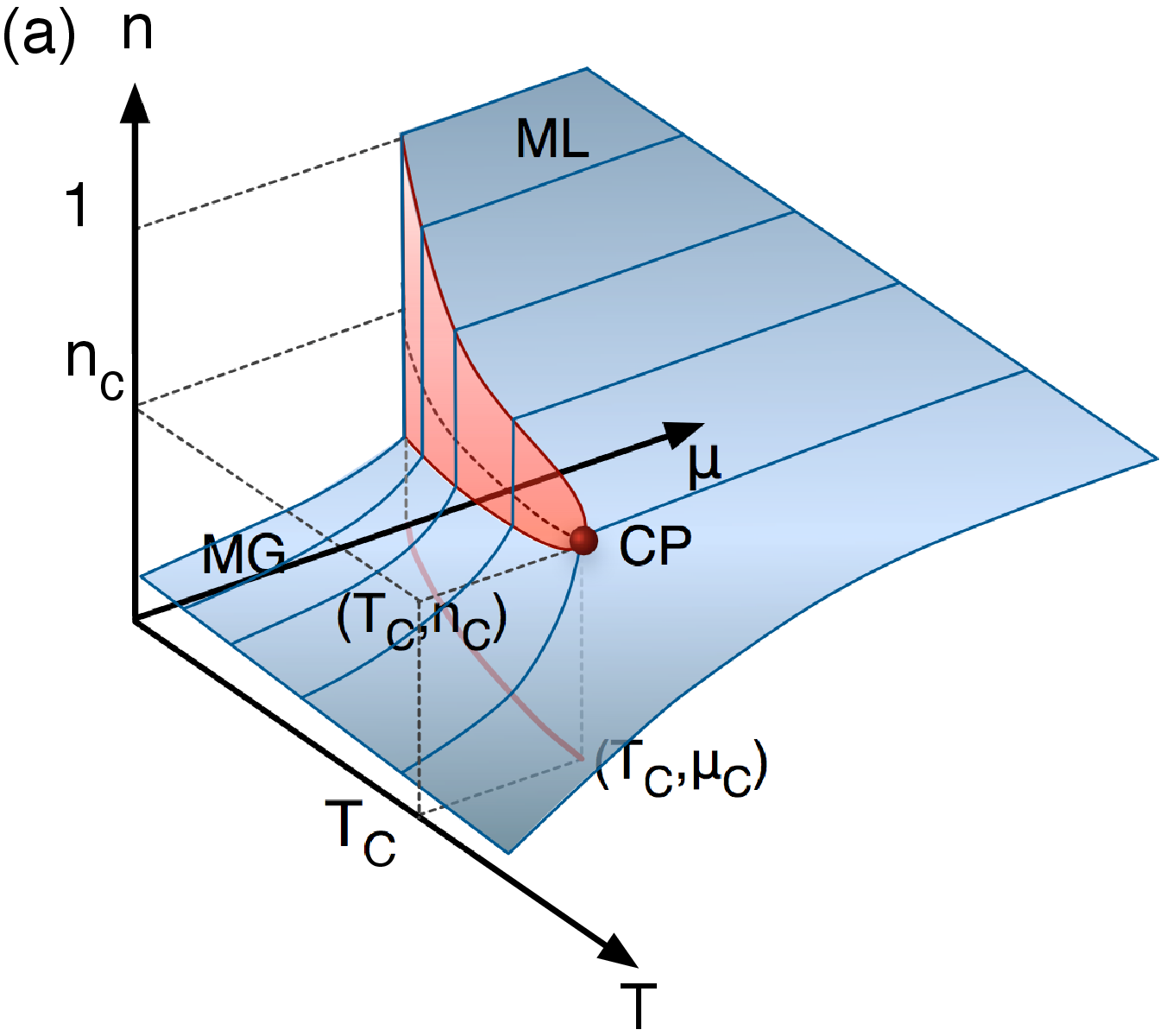}
\includegraphics[height=0.40\textwidth]{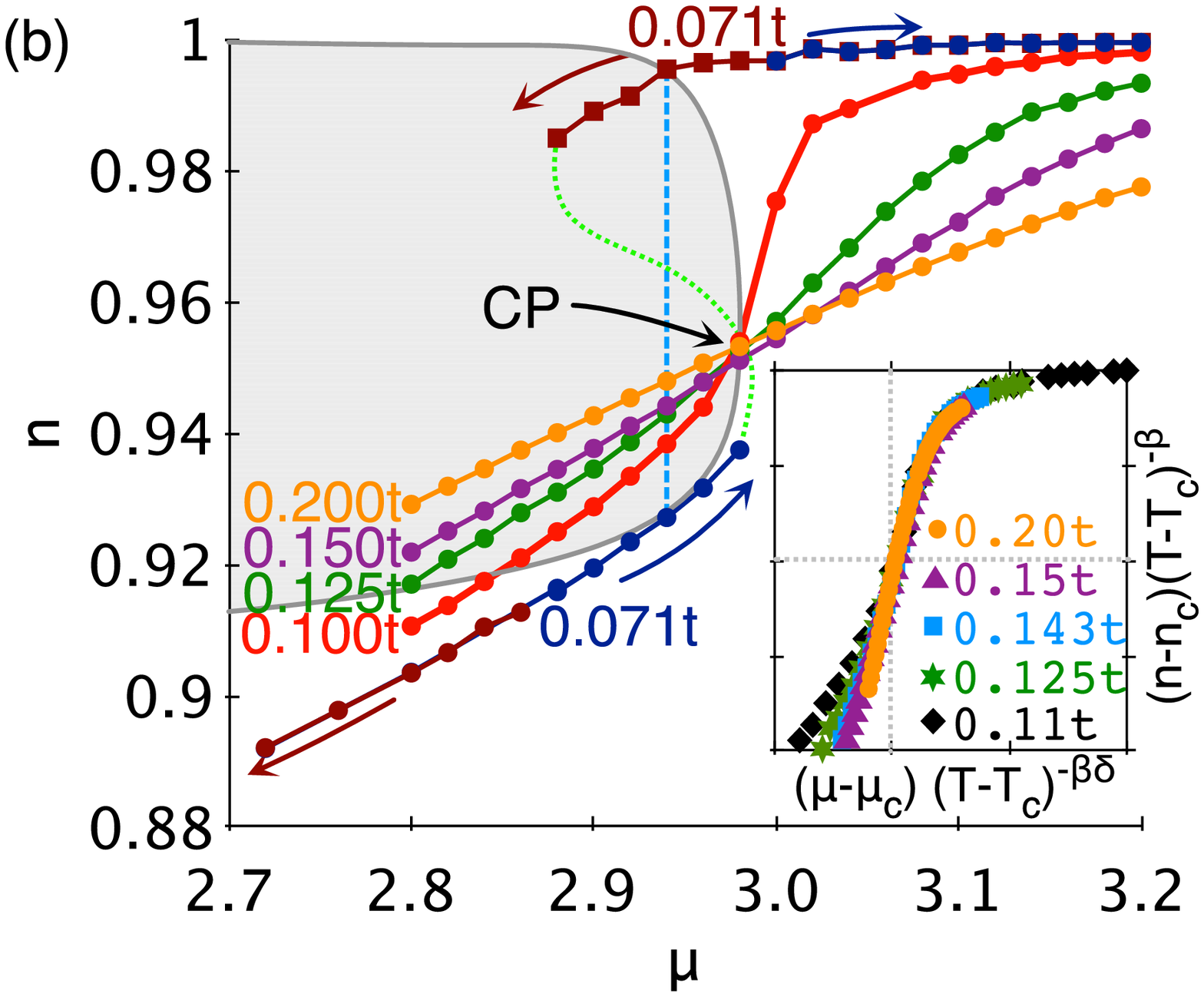}
\caption{{\bf a)} The schematic phase diagram in the presence of charge separation. This phase 
diagram describes the transition between two states labeled Mott Liquid (ML) and Mott Gas (MG) 
as a function of temperature, $T$, chemical potential, $\mu$, and filling, $n$. The red surface 
represents the coexistence region, which terminates in a critical point (CP). As we go around the 
critical point the state changes smoothly from ML to MG. Along the first order transition 
line and for a fixed $T$ and $\mu$, the filling has two values.  {\bf b)} Filling as a function 
of chemical potential for several temperatures in the vicinity of the charge separation critical 
point. The number next to each curve represents the temperature. The coexisting phases are an 
incompressible Mott liquid at 
$n\approx1$ and a compressible Mott gas at $n\approx0.93$. The critical temperature is $T_c=0.1t$. 
The blue dashed line represents the surface of metastability which is not accessible within the DCA. 
The green dotted line represents the isothermal of the metastable state inside the phase
coexistence region (gray zone). At the critical point the isothermals for $T>T_c$ cross. The inset shows 
the scaling curve
$(n-n_c)(T-T_c)^{-\beta}$ vs $(\mu-\mu_c)(T-T_c)^{-\beta\delta}$ in arbitrary units for $\mu_c=3t$, $n_c=0.96$, $T_c=0.1t$. 
The scaling exponents, $\beta=0.10 \pm 0.05$ and $\beta \delta \sim 1$, are roughly consistent with the Ising universality class.
\label{fig:Phase-separation}}
\end{center}
\end{figure}

Our findings suggest that the Hubbard model displays a phase diagram 
similar to the one for the gas-liquid transition with Mott liquid (ML) 
and Mott Gas (MG) regions.
Fig.~\ref{fig:Phase-separation}(a) shows a possible phase diagram for the Hubbard 
model as a function of $T$, $|\mu|$, and $n$. 
The red-colored surface is a schematic of the region where the Mott liquid and Mott gas 
states, characterized by different densities, coexist for $T<T_c$. The critical point is located at 
temperature $T_c$, filling $n_c$, and chemical potential $\mu_c$. One can go from one state 
to the other either smoothly, by avoiding the phase separation region, or through a first-order 
transition by crossing it. Right on the phase separation region, the density has two values 
for a given value of $\mu$ and $T$. 

\begin{figure}
\begin{center}
\includegraphics[width=0.95\textwidth]{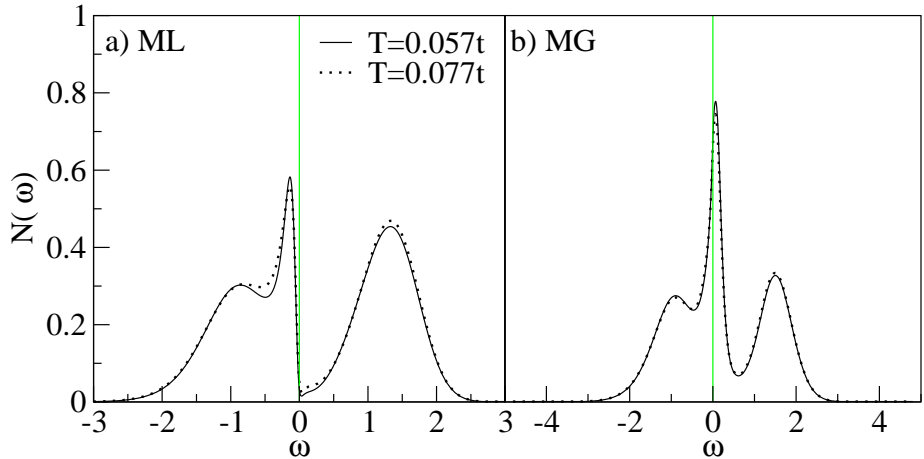}
\caption{The density of states of the {\bf a)} Mott liquid and {\bf b)} Mott gas states at 
$T=0.077t$ (dotted line) and $T=0.057$ (solid line). The Mott liquid is an incompressible
insulator with a pseudogap while the Mott gas is weakly compressible with a Fermi
liquid peak in the DOS.
\label{fig:dos_mg_ml} }
\end{center}
\end{figure}

Macridin {\em et al.} \citep{macridin_phases} provided compelling evidence of phase separation 
in the case of the generalized Hubbard model (Eq.~(\ref{eq:gen_hubbard})) with positive 
next-near-neighbor hopping $t^{\prime}=0.3\,t$ and $U=8t$. Using the DCA in a $N_c=8$ cluster 
with HFQMC as the cluster solver, they showed that below a critical temperature $T_{c}\sim 0.1t$
a first order transition occurs, which is identified by a hysteresis in the $n$ versus $\mu$ curve 
for $T<T_c$.  As shown in Fig.~\ref{fig:Phase-separation}(b) with more precise data obtained using 
DQMC as the cluster solver, the hysteresis is between two states of different filling, the Mott 
liquid at half filling and the Mott gas at a filling of about 0.93 for $T=0.071t$. The Mott liquid 
is incompressible and insulating. Its compressibility, which is the slope of the filling vs $\mu$ 
curve in the high filling side of the hysteresis curve, is small and decreases with temperature. 
Also the density of states of the ML phase, shown in Fig.~\ref{fig:dos_mg_ml}(a), exhibits a gap 
as expected for an insulator. On the other hand, the Mott gas is compressible and metallic; the 
density of states is finite at the chemical potential ($\mu=\omega=0$), as displayed in 
Fig.~\ref{fig:dos_mg_ml}(b).

The analogy to the well-known phase diagram of a liquid-gas mixture, such as water 
and steam, is useful to understand this phase transition.  At low temperatures, there 
is a region in the pressure-volume phase diagram in which water and steam coexist for
a range of pressures.  As the temperature is increased, the region of coexistence
contracts and finally terminates at a critical point where the compressibility
diverges.  In the pressure-temperature phase diagram, this region of coexistence
becomes a line of first order transitions which terminates at a second order 
point where the water and gas become indistinguishable and the compressibility 
diverges.  Since the line terminates, it is possible for the system to evolve 
adiabatically from steam to water without crossing a phase transition line; 
therefore, the steam and water must have the same symmetry.  

In the Mott liquid and Mott gas system the chemical potential $\mu$ replaces the pressure and the 
density $n$ replaces the volume of the water-gas mixture.  
Because the order parameter separating the 
ML from the MG, the density $n$, does not have a continuous symmetry, order may occur at finite 
temperatures, and the ML-MG transition will most likely be in the Ising or lattice gas universality 
class.  Within this context, one may then understand the hysteresis of 
Fig.~\ref{fig:Phase-separation}(b). 
The solid lines are isotherms which show how the system evolves with increasing density.  
At the temperature $T=T_c$, the compressibility diverges at the critical filling. 
As the temperature is lowered further, there is a region where the ML and MG coexist. Inside
this region the isothermals contain unphysical regions of negative compressibility (dashed 
green line in Fig.~\ref{fig:Phase-separation}(b)) along with metastable regions of positive 
compressibility. The metastable branch of the isothermal in the vicinity of the ML is a "supercooled"
ML, whereas the one in the vicinity of the MG is a "superheated" MG. The translational invariance
of DCA along with the stabilizing effect of the mean-field host enable access to those metastable 
states. However the real physical system will phase separate and the two phases will co-exist in 
equilibrium (dotted blue line in Fig.~\ref{fig:Phase-separation}(b)).

We can sketch the phase diagram in the $T-\mu$ plane using  the analogy to the
water-steam mixture.  The most generally applicable rule governing the shape 
of phase diagrams was established by Gibbs.  For a system of $c$ conserved 
components and $f$ phases,  the Gibbs constraint is give by the relation $\Phi=c-f +2$ where 
$\Phi$ is the number of independent variables needed to specify the state of every phase.  
In this case, as in the water-steam system, the number of components $c=1$, since the particle number
is conserved.
At a location in the phase diagram 
where only one phase exists, $\Phi=1-1+2 =2$, so there are two independent variables, 
and the phase diagram is a surface on the $\mu$, $T$ and $n$ three-dimensional space.
There will be places in the phase diagram where two phases 
exist simultaneously, then $\Phi=1-2+2=1$, implying that two phases co-exist only along lines in the 
phase diagram.
At the lines in the $T-\mu$ plane where two phases coexist, 
$n$ is also determined for each phase, but its value can be different. That is a line of first order 
transitions.

\begin{figure}
\begin{center}
\includegraphics[width=0.95\textwidth]{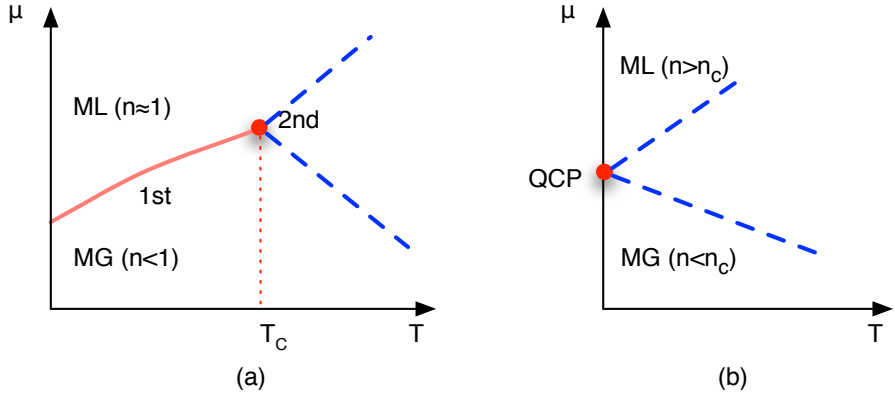}
\caption{(a) The chemical potential-temperature phase diagram of the ML and MG mixture for $t'>0$.
The ML and MG coexist on a line of first order transitions with positive slope. Since ML and MG 
have the same symmetry this line can terminate in a second order critical point. The blue dashed 
lines define the boundaries of the supercritical region where the ML and MG cannot be distinguished.  
Outside this region either the ML or MG character dominates. 
(b) The chemical potential-temperature phase diagram for $t'\rightarrow0$.  The first-order line is 
absent but supercritical region remains as a quantum critical region. In the Hubbard model the lines 
$T*$ and $T_X$ 
(Fig.~\ref{fig:Quasiparticle-weight-raja}(b)) define the boundaries of this region.
\label{fig:T-mu_classical}}
\end{center}
\end{figure}

Additional information about the lines of  first order transitions is obtained 
from the Clapeyron's equation.  The Gibbs free energy $G= E -TS - \mu N$, and $dG = -SdT -Nd\mu$,
must be the same for the coexisting phases on a line.   If we label the two phases $1$ and $2$, 
then
\begin{equation}
(S_1-S_2)dT = -(N_1-N_2)d\mu \,.
\end{equation}
If we identify the latent heat $L = (S_1-S_2)T$, then $d\mu/dT = -L/(T \Delta n)$ represents the slope 
of the first order transition line.  Since the latent heat $L$ of going  from ML to MG is 
positive, but $dn$ is negative, the slope $d\mu/dT$ of the line of first order transitions 
is positive.  

Above the critical point terminating the ML-MG transition,
the system displays supercritical behavior in a region where the gas and the liquid 
cannot be distinguished thermodynamically (c.f. Fig.~\ref{fig:T-mu_classical}).  It is possible 
for the system to evolve adiabatically through a counterclockwise path from deep in the MG 
region, through the supercritical region, into the ML region.  At the lower edge of the 
supercritical region, the system loses the Fermi-liquid character of the MG, and at the upper 
edge, it begins to acquire the pseudogap character of the ML.   

\begin{figure}
\includegraphics[width=0.5\textwidth]{ehsan_nvsmu_Nc16B_U1\lyxdot 5_tp0\lyxdot 0-0\lyxdot 3}
\includegraphics[width=0.5\textwidth]{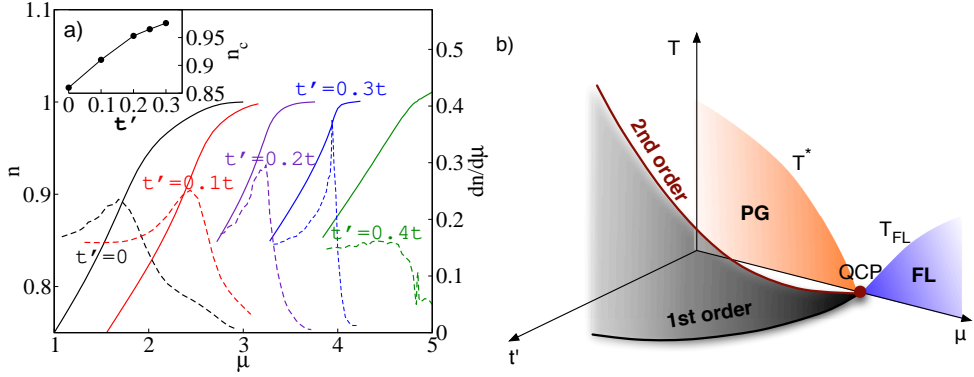}
\caption{(a) Filling, $n$,  vs.\ chemical potential, $\mu$, for $T=0.077t$, $N_c=16$, $U=6t$ and 
various $t^{\prime}$ is shown in solid lines and the compressibility $\displaystyle \frac{dn}{d\mu}$ 
in dashed lines. A critical filling, identified by the peak in the compressibility appears at higher 
temperatures and fillings as $t^\prime$ is increased. The inset shows the $t^\prime$ dependence of 
the critical filling, $n_c$. (b) Schematic phase diagram of the Hubbard model in the $\mu$, $t^\prime$ 
and $T$ space (neglecting superconductivity). The classical critical point turns asymptotically
into a quantum critical point as $t^\prime \rightarrow 0$.
\label{fig:mu-vs-n-ehsan}\label{fig:mu-tp-T-phase-diagram}}
\end{figure}

Let us discuss now 
how this phase separation, which occurs
at finite temperature, is related to quantum criticality.
The key parameter is the next-nearest-neighbor hoping, $t^{\prime}$.
For $t^{\prime}=0$ there is no evidence for phase separation at finite $T$, but
such a phase separation occurs for positive $t^{\prime}$. Khatami {\em et al.}
~\citep{khatami_criticality} performed a systematic analysis of the phase
diagram of the extended Hubbard model as a function of $t^{\prime}$. 
As shown in Fig. \ref{fig:mu-vs-n-ehsan} (a)
the compressibility, $\chi_{c}=dn/d\mu$, exhibits a peak for all positive
$t^{\prime}$ at a critical filling that depends on $t^{\prime}$.
The width of the peak measures the distance from the critical temperature:
the sharper the peak the closer to $T_{c}$ the employed temperature
is. We see that the critical temperature increases with $t^{\prime}$
and it starts from $T_{c}=0$ at $t^{\prime}=0$.
These results point to the phase diagram of Fig.~\ref{fig:mu-tp-T-phase-diagram} (b).
At a positive $t^{\prime}$ a charge separation occurs at temperatures
$T<T_{c}(t^{\prime})$ and at a critical filling $n_{c}(t^{\prime})$
between an incompressible and insulating Mott liquid and a compressible
metallic Mott gas. Right at $T_{c}$, there is a terminating second-order 
critical point. By decreasing $t^{\prime}$ this second-order
critical point is pushed down to lower temperatures. Presumably the
line of second-order critical points terminates at the QCP.

Such a scenario constitutes a new path to quantum criticality as it is closely tied to charge 
fluctuations rather than spin fluctuations.  However, numerous simulations suggest that a finite 
positive $t'$ enhances antiferromagnetic correlations, and since phase separation is only present 
for $t'>0$ it suggests that it is driven by strong spin correlations.  In addition, previous simulations 
incorporating Holstein phonons to the Hubbard model found that phonons also enhance the phase 
separation instability~\citep{macridin_isotope}.  As $t'/t \to 0$ (and the electron-phonon coupling 
vanishes), the phase separation critical point approaches zero temperature becoming a QCP.  Here, the 
first-order behavior is absent from the  phase diagram (Fig.~\ref{fig:Quasiparticle-weight-raja}(b)) 
leaving only the adiabatic path from the ML to the MG, which 
passes through the supercritical region, which is  now  the quantum critical (QC) region.  
The crossover scale $T_X$ and the pseudogap scale $T^*$  are now understood as the boundaries of 
the QC region.  As we cross  the line of  $T^*$ from the QC region into the ML 
region, the characteristics of the ML become apparent, including the pseudogap in the DOS and the 
insulating behavior.  As we cross the line of $T_X$ from the QC region into the MG, the 
characteristics 
of the MG become apparent, including Fermi liquid formation.

Those calculations certainly do not elucidate the nature of the Mott
liquid and Mott gas states in real materials. The long-ranged nature of the 
Coulomb interaction prevents true charge separated states, but the phase separation
we observe may also correspond to other charge instabilities,
such as stripes or checkerboard patterns. To distinguish between different
charge instabilities, systematic calculations in much larger clusters
are necessary which are not practical at the moment. However, whatever
the type of order, those calculations provide convincing evidence
for the existence of a first-order transition at low temperatures.
Such a transition is similar to the liquid-gas 
or the ferromagnetic transition and its phase diagram 
would look like Fig. 
\ref{fig:Phase-separation}(a): a first order line of coexistence which
terminates at a critical point at a critical temperature $T_{c}$ and a 
critical filling $n_{c}$. 

\section{Conclusions}

The presence of a QCP at finite filling in the cuprate phase diagram
is a topic of active theoretical and experimental research. Quantum cluster
methods are able to shed some light in this phase diagram. By studying
single-particle quantities for $t^{\prime}=0$, such as the spectral function 
and the entropy, it can be shown that a Fermi-liquid region at low filling 
and the pseudogap region at higher filling have different spectral signatures, 
and are connected through an intermediate "marginal Fermi-liquid" region of 
maximal entropy. Due to limitations of quantum Monte Carlo, the ground state
and quantum criticality are not accessible. We also neglect the superconducting
phase transition. The connection with 
quantum criticality is established by switching on $t^{\prime}$. For positive
$t^{\prime}$ a classical critical point emerges at finite temperature $T_c$, which
increases with $t^\prime$. We note that $t^\prime$ is not the
only control parameter that may be able to tune the critical point to finite
temperatures, but other parameters, such as phonon coupling, may have the same
effect. The phase diagram around the critical point is similar to that of the
gas-liquid transition, where the incompressible Mott liquid and the compressible
Mott gas are the coexisting phases. The strange metal region in this context 
may be viewed as the supercritical region lying in the vicinity of the critical 
point. Within the scenario we presented, the pseudogap region is not characterized
by an order parameter, rather it must have the same symmetry as the Fermi-liquid  
and the marginal Fermi-liquid, since these regions are connected by
an  adiabatic path in the $T-\mu$ phase diagram. Further investigation is necessary 
to fully characterize the pseudogap region, and also to investigate the connection
of those results with other scenarios of quantum criticality.

\section{Acknowledgements}
We would like to thank R. Gass, S. Kivelson, D. J. Scalapino, A. M. Tremblay, 
C. Varma, M. Vojt, S. R. White, J. Zaanen and  for useful discussion that helped
during the development of the presented work. This research was supported by NSF DMR-0706379, 
DOE CMSN DE-FG02-08ER46540, and by the DOE SciDAC grant DE-FC02-06ER25792. This research 
used resources of the National Center for Computational Sciences at Oak Ridge National 
Laboratory, which is supported by the Office of Science of the U.S. Department of Energy 
under Contract No. DE-AC05-00OR22725.

\bibliographystyle{rspublicnat}

\end{document}